\documentclass[pra,aps,
amssymb,amsmath,amsmath,showpacs,reprint,twocolumn]{revtex4-1}
\usepackage{color}
\usepackage{graphicx}
\usepackage{epstopdf}
\usepackage{natbib}
\usepackage{wasysym}
\usepackage{hyperref}
\usepackage{dcolumn}

\hypersetup{%
   pdfpagemode=None, 
   pdfstartpage=1,
   pdfstartview=FitH,
   pdfmenubar=true,
   pdftoolbar=true,
   colorlinks = true,
   linkcolor=blue,
   citecolor=blue,
   bookmarksopen=false
}

\newcommand{\Fkt}[1]{\,\mathsf {#1}}
\def\openone{\leavevmode\hbox{\small1\kern-3.3pt\normalsize1}}

\ifx\Tr\renewcommand{\Tr}{\Fkt{Tr}} 
\else\newcommand{\Tr}{\Fkt{Tr}}
\fi

\usepackage{booktabs}
\usepackage{graphicx}
\usepackage{amsmath}
\usepackage{indentfirst}
 \newcommand{\vecr}{\mathbf{r}}
 \newcommand{\half}{\frac{1}{2}}
 \newcommand{\rb}{r_>}
 \newcommand{\rs}{r_<}

\begin{document}
\title{Calculation of Araki-Sucher correction for many-electron systems}

\author{\sc Justyna G. Balcerzak}
\author{\sc Micha\l\ Lesiuk}
\email{e-mail: lesiuk@tiger.chem.uw.edu.pl}
\author{\sc Robert Moszynski}
\affiliation{\sl Faculty of Chemistry, University of Warsaw\\
Pasteura 1, 02-093 Warsaw, Poland}
\date{\today}

\begin{abstract}
In this paper we consider the evaluation of the Araki-Sucher correction for arbitrary many-electron atomic and molecular
systems. This contribution appears in the leading order quantum electrodynamics corrections to the energy of a bound
state. The conventional one-electron basis set of Gaussian-type orbitals (GTOs) is adopted; this leads to
two-electron matrix elements which are evaluated with help of generalised the McMurchie-Davidson scheme. We also
consider the convergence of the results towards the complete basis set. A rigorous analytic result for the convergence
rate is obtained and verified by comparing with independent numerical values for the helium atom. Finally, we present a
selection of numerical examples and compare our results with the available reference data for small systems. In contrast
with other methods used for the evaluation of the Araki-Sucher correction, our method is not restricted to few-electron
atoms or molecules. This is illustrated by calculations for several many-electron atoms and molecules.
\end{abstract}

\maketitle

\section{Introduction}
\label{sec:intro}

In the past few decades there has been remarkable a progress in the many-body electronic structure theory. This has
allowed to treat large systems of chemical or biological significance containing hundreds of electrons and, at the same
time, obtain very accurate results for small systems which are intensively studied spectroscopically. Introduction
of general explicitly correlated methods \cite{hattig12,kong12,tenno12}, reliable extrapolation techniques
\cite{helgaker97,halkier98,halkier99a,halkier99b,halkier99c,truhlar98}, general coupled cluster theories
\cite{kallay01,hirata03}, and new or improved one-electron basis sets
\cite{dunning89,woon95,peterson02,kemmish14,kemmish15a,kemmish15b,lesiuk12,lesiuk14a,lesiuk14b,lesiuk15,lesiuk16} made
the so-called spectroscopic accuracy (few cm$^{-1}$ or less) achievable for many small molecules.

However, as the accuracy standards of routine calculations are tightened up one encounters new challenges. One of these
challenges is the necessity to include corrections due to finite mass of the nuclei \cite{born56},
relativistic, and quantum electrodynamic (QED) effects \cite{bethe75}, and possibly finite nuclear size
\cite{vissher97}. The former two have been subjects to many
studies in the past decades, see Refs. \cite{kolos64,wolniewicz93,kutz97,valeev03,gauss06,pykko78,dyall07,reiher09} and
references therein. However, systematic studies of importance of the QED effects in atoms and molecules are
scarce and have
begun relatively recently \cite{beier05,kozhedub08,pasteka17}. High accuracy of the \emph{ab initio} calculations and
reliability of the theoretical predictions is of prime importance, e.g., in the field of ultracold
molecules. This is best illustrated by the papers of McGuyer et al. \cite{mcguyer15} devoted to observation of the
subradiant states of Sr$_2$ or by McDonald and collaborators \cite{mcdonald16} where photodissociation of ultracold
molecules was studied. Notably, the importance of QED effects has also been realised in the first-principles studies of
He$_2$ for the purposes of metrology \cite{jeziorski10,jeziorski12,jeziorski17}.

QED is definitely one of the most successful theories in physics, with calculation of the anomalous magnetic moment of
the electron being the prime example \cite{kinoshita06,hagiwara07,aoyama12}. However, applications to the bound states,
e.g., with a strong Coulomb field are marred with problems. Two physical phenomena, electron self-energy and
vacuum polarisation, giving rise to the Lamb shift \cite{bethe75} are difficult to include in the standard many-body
theories. For moderate and large $Z$ approaches based on the Uehling potential \cite{uehling35} with optional
corrections \cite{wichmann56,blomqvist72}, scaling of the hydrogen-like values \cite{johnson85,mohr92}, effective
potentials of Shabaev et al. \cite{shabaev02,shabaev03}, multiple commutator approach by Labzowsky and Goidenko
\cite{labzowski97,labzowski99}, and effective Hamiltonians of Flambaum and Ginges \cite{flambaum05} were used with a
considerable success. 

However, for small and moderate $Z$ the most theoretically consistent approach is the nonrelativistic QED
theory (NRQED) proposed by Caswell and Lepage \cite{caswell86} and further developed and extended by
Pachucki \cite{pachucki93,pachucki98,pachucki04a,pachucki05}. This method relies on the expansion of the exact energy in
power
series of the fine structure constant, $\alpha$. The coefficients of the expansion are evaluated as expectation values
of an effective Hamiltonian with the nonrelativistic wavefunction. Thus, the zeroth-order term is simply the
nonrelativistic energy, the first-order term is zero, the second-order contributions are expectation values of the
Breit-Pauli Hamiltonian (the relativistic corrections).

NRQED has been successfully applied to numerous few-electron atomic and molecular systems. Obvious applications are the
one-electron systems such as hydrogen-like atoms (see Ref. \cite{sapirstein90} for a comprehensive review)
and hydrogen molecular ion
\cite{bishop81,bukowski92,korobov04,korobov06,korobov13}. Beyond that, very accurate results are available for the
helium atom
\cite{schwartz61,yan98,korobov99,drake99,cencek01,pachucki00a,pachucki00b,pachucki06a,pachucki06c,lach04,pachucki17a,
pachucki17b}, hydrogen molecule
\cite{garcia66,bishop78,wolniewicz95,kpisz09,komasa11,puchalski16}, and their isotopomers. Remarkably,
corrections of the order $\alpha^4$ have been derived and evaluated recently
\cite{pachucki06a,pachucki06b,puchalski16,pachucki10b}. Other examples are
lithium \cite{yan02,pachucki03,puchalski13a} and beryllium atoms \cite{puchalski13b} with the corresponding ions
\cite{pachucki06b},
and helium dimer \cite{jeziorski10,jeziorski12,cencek12}. In all these examples very accurate agreement with the
experimental data has been obtained
which confirms validity and applicability of NRQED to light molecular and atomic systems. However, all presently
available rigorous methods for calculation of the NRQED corrections are inherently limited to few-body systems and
cannot be straightforwardly extended to larger ones.

In the framework of NRQED, the leading (pure) QED corrections are of the order $\alpha^3$ and $\alpha^3
\ln \alpha$. For a singlet atomic or molecular state one has \cite{araki57,sucher58}
\begin{align}
\label{qed}
\begin{split}
E^{(3)} &= \frac{8\alpha}{3\pi}\left(\frac{19}{30}-2\ln \alpha - \ln k_0\right)\langle D_1 \rangle \\
&+ \frac{\alpha}{\pi} \left(\frac{164}{15}+\frac{14}{3}\ln \alpha\right)\langle D_2 \rangle + \langle
H_{AS} \rangle,
\end{split}
\end{align}
where $\ln k_0$ is the so-called Bethe logarithm \cite{bethe75,schwartz61}; $\langle D_1 \rangle$ and $\langle D_2
\rangle$ are the one- and two-electron Darwin terms
\begin{align}
\langle D_1 \rangle &= \frac{\pi}{2}\,\alpha^2\sum_a Z_a \langle \sum_i \delta(\textbf{r}_{ia})\rangle,\\
\langle D_2 \rangle &= \pi\,\alpha^2\,\langle\sum_{i>j}\delta(\textbf{r}_{ij})\rangle,
\end{align}
where $Z_a$ are the nuclear charges, and $\delta (\vecr)$ is the three-dimensional Dirac delta distribution. Throughout
the paper, we use letters $i,j,\ldots$ and $a,b,\ldots$ to denote summations over electrons and nuclei, respectively.
The last term in Eq. (\ref{qed}) is the Araki-Sucher correction
\begin{align}
\label{araki1}
\langle H_{AS} \rangle = -\frac{7\alpha^3}{6\pi} \langle \sum_{i>j} \hat{P}\left(r_{ij}^{-3}\right)\rangle,
\end{align}
where the regularised distribution in the brackets is defined by the following formulae
\begin{align}
\label{AS1}
 \hat{P}(r_{ij}^{-3}) = \lim_{a \rightarrow 0} \hat{P}_a (r_{ij}^{-3}),
\end{align}
and
\begin{align}
\label{AS2}
 \hat{P}_a (r_{ij}^{-3}) = \theta (r_{ij} - a) \, r_{ij}^{-3} + 4\pi 
 \left(  \gamma + \ln a \right) \delta (\vecr_{ij}),
\end{align}
where $\gamma$ is the Euler–Mascheroni constant, and $\theta(x)$ is the Heaviside step function in the usual convention.

In evaluation of the QED corrections for many-electron atoms and molecules, two quantities present in Eq.
(\ref{qed}) are the major source of difficulties. The first one is the Bethe logarithm and the second is the
Araki-Sucher correction. In this paper we are concerned with the latter quantity; evaluation of the Bethe logarithm will
be considered in subsequent papers. Let us point out that the Araki-Sucher term is not necessarily the largest of the
$\alpha^3$ QED corrections. In fact, the one-electron terms typically dominate in Eq. (\ref{qed}). However, the
relative importance of the Araki-Sucher correction is expected to increase for heavier atoms similarly as for the
two-electron terms of the Breit-Pauli Hamiltonian. Moreover, for polyatomic systems the Araki-Sucher correction
possesses an usual $R^{-3}$ asymptotics for large interatomic distances, $R$. As a result, it decays much less rapidly
than the other components of the potential energy curve \cite{pachucki05b} and its importance is substantial for large
$R$.

From the point of view of the many-body electronic structure theory Eq. (\ref{araki1}) is an ordinary expectation value
of a two-electron operator. Provided that the corresponding matrix elements are available, evaluation of such
expectation values by using the coupled cluster (CC) \cite{bartlett07} or configuration interaction (CI) wavefunctions
is a standard task \cite{salter89,hald03,coriani04,jeziorski93,moszynski05,korona06,tucholska14,tucholska17}. Therefore,
in this work we are concerned with evaluation of the matrix
elements (i.e., two-electron integrals) of the Araki-Sucher distribution in the Gaussian-type orbitals (GTOs)
basis \cite{boys50}. Importantly, the proposed method can be applied to an arbitrary molecule and is not limited to
few-electron systems.

Throughout the paper, we follow Ref. \cite{stegun72} in definitions of all special and elementary functions.

\section{Calculation of the matrix elements}
\label{sec:integrals}

In this section we consider evaluation of matrix elements necessary to calculate the Araki-Sucher correction for
many-electron atomic and molecular systems. We adopt the usual Gaussian-type orbitals (GTOs) in the Cartesian
representation \cite{boys50} as one-electron basis set
\begin{equation}
\label{gto}
 \phi_a(\vecr_A) = x_A^i\, y_A^k\, z_A^m\, e^{-a r_A^2},
\end{equation}
where $\mathbf{A} = ( A_x, A_y, A_z )$ is a vector specifying location of the orbital, $x_A = x - A_x$ and similarly 
for the remaining coordinates. For brevity, we omit the normalisation constant in the definition (\ref{gto}). However,
normalised orbitals are used in all calculations described further in the paper.

Evaluation of the Araki-Sucher correction from the many-electron coupled cluster wavefunction within the basis set
(\ref{gto}) requires the following two-electron matrix elements
\begin{align}
\label{elemas}
\begin{split}
\left( ab | cd \right) =  \iint \text{d}\vecr_1\,\text{d}\vecr_2\,
\phi_a(\vecr_{1A})\,\phi_b(\vecr_{1B})\\
\times\,\hat{P}(r_{12}^{-3})\, \phi_c(\vecr_{2C})\,\phi_d(\vecr_{2D}).
\end{split}
\end{align}
The scheme presented further in the paper relies on the McMurchie-Davidson method \cite{mcmurchie78,helgaker92}. This
method was first introduced in the context of the standard two-electron repulsion integrals and various one-electron
integrals necessary for calculation of the molecular properties. Later, it was extended to handle integral derivatives
and more involved two-electron integrals found in the so-called explicitly correlated methods \cite{samson02,sila15}.
While some other methods of calculation of the usual electron repulsion integrals are more computationally efficient
(cf. Ref. \cite{gill94}) than the McMurchie-Davidson scheme, the latter is much simpler to implement and extend to more
complicated integrals. This was the main motivation for its use in the present context.

\subsection{Generalised McMurchie-Davidson scheme}

The backbone of the McMurchie-Davidson scheme is the so-called Gauss-Hermite function, $\Lambda_t(x;a)$, defined formally as
\begin{align}
\label{hermite}
 \Lambda_t(x_A;a) \exp(-a x_A^2) = \partial_{A_x}^t\exp(-a x_A^2).
\end{align}
Clearly, the functions $\Lambda_t(x;a)$ are closely related (by scaling) with the well-known Hermite polynomials.
It is also straightforward to prove the following relation
\begin{align}
\label{trans}
 x_A^i x_B^j\,e^{-a x_A^2}\,e^{-b x_B^2} = e^{-p x_P^2} \sum_{t=0}^{i+j} E_t^{i j}\Lambda_t(x_P;p),
\end{align}
where $p=a+b$ and $\mathbf{P} = \frac{a\mathbf{A} + b\mathbf{B}}{p}$. The coefficients $E_t^{i j}$ can be calculated
with convenient recursion relations \cite{mcmurchie78,helgaker92}.

With the help of Eqs. (\ref{hermite}) and (\ref{trans}) one can show that the product of two off-centred GTOs can be
written as
\begin{align}
\label{gtoprod}
\begin{split}
\phi_a(\vecr_A)\phi_b(\vecr_B) &= \sum_{t=0}^{i+j} E_t^{ij} \sum_{u=0}^{k+l} E_u^{kl} \sum_{v=0}^{m+n} E_u^{mn} \\
&\times \partial_{P_x}^t\, \partial_{P_y}^u\, \partial_{P_z}^v\, \exp(-p r_P^2).
\end{split}
\end{align}
Returning to the initial integrals (\ref{elemas}) and use Eq. (\ref{gtoprod}) for both orbital products. This leads
to
\begin{align}
\label{abcd}
\begin{split}
 \left(ab|cd \right) &= \sum_{t=0}^{i+j} E_t^{ij} \sum_{u=0}^{k+l} E_u^{kl} \sum_{v=0}^{m+n} E_u^{mn}\,(-1)^{t+u+v}\\
 &\times\sum_{t'=0}^{i'+j'} E_{t'}^{i'j'} \sum_{u'=0}^{k'+l'} E_{u'}^{k'l'} \sum_{v'=0}^{m'+n'} E_{u'}^{m'n'}\\
 &\times R^{t+t',u+u', v+v'}, 
\end{split}
\end{align}
 where 
\begin{align}
\label{R}
 R^{tuv} = \partial_{Q_x}^t\,\partial_{Q_y}^u\,\partial_{Q_z}^v\, B,
\end{align}
and $B$ is the so-called \emph{basic integral} defined as $B=\lim_{a \rightarrow 0}B_a$ with
\begin{align}
\label{basic1}
 B_a = \iint \text{d}\vecr_1\,\text{d}\vecr_2\, \exp(-p r_{1P}^2)\,\hat{P}_a(r_{12}^{-3})\,\exp(-q r_{2Q}^2).
\end{align}
Note that differentiation with respect to the coordinates of $\mathbf{P}$ in Eq. (\ref{abcd})
has been replaced by differentiation with respect to the corresponding components of $\mathbf{Q}$. This is valid
because the basic integral is dependent only on the length of $\mathbf{P}-\mathbf{Q}$ but not on the individual
components.

The biggest inconvenience connected with Eq. (\ref{R}) is the necessity to differentiate with respect to Cartesian
coordinates. In the original treatment of McMurchie and Davidson (concerning the standard electron repulsion integrals)
a four-dimensional recursion relation was introduced to resolve this issue \cite{mcmurchie78,helgaker92}. This approach
is difficult to generalise to other basic integrals and typically requires a separate treatment in each case. In a
recent paper we proposed a different strategy based on the following expression \cite{sila15}
\begin{align}
 \label{trans0}
 x^t y^u z^v = \sum_{l=0}^{l_{max}} \sum_{m=-l}^{l} c^{lm}_{tuv}\, r^{l_{max}-l}\, Z_{lm}(\vecr),
\end{align}
where $l_{max} = t+u+v$,
relating Cartesian coordinates with the real solid spherical harmonics, $Z_{lm}(\vecr) = r^l\, Y_{lm}(\hat{r})$ (note
that the Racah normalisation is not adopted here). The numerical coefficients $c^{lm}_{tuv}$ can be precalculated and
stored in memory as a look-up table (cf. the work of Schlegel \cite{schlegel95}). In analogy, the differentials present
in Eq. (\ref{R}) are rewritten as
\begin{align}
 \label{trans1}
 \partial_{Q_x}^t\,\partial_{Q_y}^u\,\partial_{Q_z}^v = \sum_{l=0}^{l_{max}} \sum_{m=-l}^{l}
c^{lm}_{tuv}\,\nabla_Q^{l_{max}-l}\, \hat{Z}_{lm}(\nabla_Q),
\end{align}
where $\nabla_Q$ is the gradient operator and $\hat{Z}_{lm}(\nabla_Q)$ are the (real) spherical harmonic gradient
operators \cite{weniger05}. Heuristically, they are obtained by taking an explicit expression for $Z_{lm}(\vecr)$ and
replacing all Cartesian coordinates with the corresponding differentials.

By the virtue of the Hobson theorem \cite{hobson92} one has
\begin{align}
 \label{tensor1}
 \hat{Z}_{lm}(\nabla_Q)g(Q) = \left[D_Q^l\, g(Q)\right] Z_{lm}(\mathbf{Q}),
\end{align}
where $D_Q=Q^{-1}\partial_Q$, for an arbitrary function $g(Q)$ dependent only on the length of the vector, $Q$. With
help of Eqs. (\ref{trans1}) and (\ref{tensor1}) one can write
 \begin{align}
 \label{trans2}
 R^{tuv} = \sum_{l=0}^{l_{max}} \sum_{m=-l}^{l}\,c^{lm}_{tuv}\,\nabla_Q^{l_{max}-l}\,\left[ \left(D_Q^l \,
B\right) Z_{lm}(\mathbf{Q}) \right].
 \end{align}
Note that the quantity in the subscript, $l_{max}-l$, is always even (otherwise the coefficients $c^{lm}_{tuv}$ vanish).
Therefore, the last step amounts to repeated action of the Laplacian on the terms in the square brackets. The final
result can be obtained by noting that the solid harmonics are eigenfunctions of the Laplace operator and by using the
obvious relationship $\nabla_Q^2 = Q^2 D_Q^2+3D_Q$ for the radial part of the integrations
\begin{align}
 \label{trans3}
 \begin{split}
 R^{tuv} &= \sum_{l=0}^{l_{max}} \sum_{m=-l}^{l}\,c^{lm}_{tuv}\,Z_{lm}(\mathbf{Q})\sum_{k=0}^{k_{max}}
 d_k^{l,k_{max}} \\
 &\times\left(D_Q^{l_{max}-k} \,B\right) Q^{l_{max}-l-2k}
 \end{split}
\end{align}
where $k_{max}=\half(l_{max}-l)$. The auxiliary coefficients, $d_{n}^{lm}$, are calculated recursively
\begin{align}
\begin{split}
 d_{n}^{lm} &= d_{n}^{l,m-1} + \left[2l+3+4(m-n)\right] d_{n-1}^{l,m-1} \\ &+ 2(m-n+1)(2l+3+2(m-n))d_{n-2}^{l,m-1}
\end{split}
\end{align}
starting with $d_0^{lm}=1$; the last term of the recursion is neglected for $n=1$. Note that the coefficients
$d_{n}^{lm}$ can also be stored as a look-up table. 

To sum up, by means of Eq. (\ref{trans3}) all integrals $R^{tuv}$ are
expressed through the derivatives of the basic integral, $D_Q^l B$. We consider evaluation of these quantities in the
next section. Let us also note in passing that to achieve an optimal efficiency during the evaluation of Eq.
(\ref{trans3})
the summations need to be carried out stepwise, paying attention to the order of the individual sums.
 
 \subsection{Basic integral and derivatives}
 
Calculation of the basic integral, given formally by the limit of Eq. (\ref{basic1}), is hampered by the troublesome
form of the Araki-Sucher distribution. It was shown in Ref. \cite{sila15} that an equivalent general formula for the
basic integral reads
\begin{align}
 B_a = e^ {-q Q^2} \iint \text{d} \vecr_1 \text{d} \vecr_2\, e^{-q r^2_2-p r^2_1}\, i_0\left(2qQr_2\right)\,
\hat{P}_a(r_{12}^{-3}),
\end{align}
where $i_0(x)=\sinh x/x$. In the present case this expression naturally splits into two parts, $B_a = B_a^{(1)} +
B_a^{(2)}$
\begin{align}
 B_a^{(1)} &= e^ {- q Q^2} \iint \text{d} \vecr_1 \text{d} \vecr_2\, 
\frac{\theta\left(r_{12}-a\right)}{r_{12}^{3}}e^{-q r^2_2-p r^2_1}\, i_0\left(2qQr_2\right), \\
 B_a^{(2)} &= 4\pi\, e^ {-q Q^2} (\gamma +\ln a)\int \text{d} \vecr\,e^{-(p+q) r^2}\, i_0\left(2qQr\right),
\end{align}
where the second formula follows directly from the properties of the Dirac delta distribution.
The first integral can be simplified by changing the coordinates to $r_1$, $r_2$, $r_{12}$ and three arbitrary angles. 
Integration over all variables apart from $r_{12}$ is elementary
\begin{align}
 B_a^{(1)} = \sqrt{\frac{\pi^5}{p+q}} \frac{1}{pq} \int_a^{\infty} \, \frac{\text{d} r}{r^2} \frac{e^{-\xi(r-Q)^2}-e^{-\xi(r+Q)^2}}{Q},
\end{align}
where $\xi = \frac{pq}{p+q}$. The next step is to expand the $Q$-dependent part of the integrand into power series
\begin{align}
\label{B_1}
 \begin{split}
  B_a^{(1)} = 2\mathcal{N}e^{-\xi Q^2} \sum_{n=0}^{\infty} \frac{(2\xi Q)^{2n}}{(2n+1)!}\int_a^{\infty} \text{d}r\, r^{2n-1}\,e^{-\xi r^2},
 \end{split}
\end{align}
where $\mathcal{N}=2\pi\,\left(\frac{\pi}{p+q}\right)^{3/2}$.
The first term of the series (corresponding to $n=0$) must be extracted and treated separately, but the remaining
integrals are straightforward. Importantly, to simplify the integration process we drop all higher-order terms in $a$
which do not contribute to the final result (once the $a\rightarrow0$ limit is taken). After integration and some
rearrangements one obtains
\begin{align}
\begin{split}
 B_a^{(1)} &= 2\mathcal{N}\,e^{-\xi Q^2}\bigg[-\frac{\gamma}{2}-\ln a-\frac{\ln \xi}{2} \\
 &+ \frac{1}{2}\sum_{n=1}^{\infty}\frac{4^n(n-1)!}{(2n+1)!}\left(\xi Q^2\right)^n\bigg]+\mathcal{O}(a).
\end{split}
\end{align}
Let us return to the second part of the basic integral, $B_a^{(2)}$. Fortunately, this integration is elementary
\begin{align} 
 B_a^{(2)} = 2\mathcal{N}e^{-\xi Q^2} \left(\gamma +\ln a\right).
\end{align}
Let us now add both contributions and take the limit $a\rightarrow 0$. The logarithmic singularities present in
$B_a^{(1)}$ and $B_a^{(2)}$ cancel out, and the result reads
\begin{align}
\label{basic2}
\begin{split}
 B = \mathcal{N}e^{-\xi Q^2} \bigg[\gamma-\ln \xi 
 + \sum_{n=1}^{\infty}\frac{4^n(n-1)!}{(2n+1)!}\left(\xi Q^2\right)^n\bigg].
\end{split}
\end{align}
One can easily prove that the infinite series present in the above expression is convergent for an arbitrary real $\xi
Q^2$. Therefore, this formula constitutes an exact analytical result. However, the rate of convergence of this series
can be expected to be very slow for large values of the parameter, greatly increasing the cost of the calculations.
Moreover, this representation does not allow for a straightforward calculation of the derivatives, $D_Q^l$. Therefore,
it is desirable to bring this expression into a more computationally convenient form.

For this sake, the series in Eq. (\ref{basic2}) is summed analytically giving the following integral representation
\begin{align}
\label{resum}
 \sum_{n=1}^{\infty}\, \frac{4^n(n-1)!}{(2n+1)!}\,x^n = \int_0^1 \text{d}t\,(1-t)^{1/2}\,\frac{e^{tx} - 1}{t}.
\end{align}
Validity of this formula can easily be verified by expanding the integrand into power series in $x$ and integrating
term-by-term. Guided by Eq. (\ref{resum}) one can introduce a more general family of functions
\begin{align}
\label{jl}
 J_l(x) = e^{-x} \int_0^1 \frac{\text{d} t}{t} \, (1-t)^{1/2} \, \left[e^{tx} (1-t)^l - 1\right].
\end{align}
Note that $J_0(x)$ directly corresponds to the result of the summation in Eq. (\ref{resum}) and $J_l(x)=-\partial_x
J_{l-1}(x)$. With the help of the newly introduced quantities the basic integral is rewritten as
\begin{align}
\label{basic3}
\begin{split}
 B = \mathcal{N} \Big[e^{-\xi Q^2}\left(\gamma-\ln \xi\right) + J_0\left(\xi Q^2\right)\Big].
\end{split}
\end{align}
Within this particular representation of the basic integral it becomes disarmingly simple to perform the required
differentiation. In fact, one can show that
\begin{align}
\label{basicder}
 D_Q^l B = \mathcal{N}(-2\xi)^l \Big[e^{-\xi Q^2}\left(\gamma-\ln \xi\right) + J_l\left(\xi Q^2\right)\Big],
\end{align}
which completes the present section. Parenthetically, we note that in the above formula the argument of $J_l$ is always
positive, i.e., $\xi Q^2>0$, despite the fact that the formal definition of these integrals given by Eq.
(\ref{jl}) is valid for an arbitrary complex-valued $x$.

At this point we would like to compare our results with some other expressions published in the literature.
An integral closely related to the basic integral $B$ was considered in Ref. \cite{kpisz09}. In fact, one can
verify that by setting $c_1=c_2=0$ in Eq. (15) of Ref. \cite{kpisz09} one obtains Eq. (\ref{basic1})
of the present work (after taking the $a\rightarrow 0$ limit). However, no results for the derivatives $D_Q^l B$ were
provided as they do not appear in the explicitly correlated Gaussian calculations. Interestingly, an alternative
integral representation of $J_0(x)$ was given in Ref. \cite{kpisz09}. In our notation
\begin{align}
\label{j0alt}
 J_0(x) = e^{-x} \int_0^x \frac{\text{d} t}{t}\left[\sqrt{\frac{\pi}{t}} \frac{e^t}{2}\,\mbox{erf}(\sqrt{t})-1 \right],
\end{align}
where $\mbox{erf}(x)$ is the error function.
One can verify that the definitions (\ref{jl}) and (\ref{j0alt}) coincide by exchanging the variables and working out
the inner integral. We have not found the above representation particularly useful in the present context, but it
provides an additional verification that our final result is correct.

 \subsection{Auxiliary integrals $J_l(x)$}

The only missing building block of the present theory is the calculation of the integrals $J_l(x)$. First, let us
specify the range of parameters ($x$ and $l$) which are of interest. The maximal value of $l$ is set to $32$ in our
program.
This allows to compute the integrals (\ref{elemas}) with the maximal value $i+k+m=8$ in the one-electron basis set, see
Eq. (\ref{gto}). This corresponds to the maximal value of the angular momentum $l=8$ ($L$-type functions) in a purely
spherical representation. The are no limitations on the value of $x\geq 0$, i.e., the code is open-ended with
respect to positive values of $x$. Below, we provide a set of procedures based mostly on the recursive relations which
allow to calculate the integrals $J_l(x)$ with accuracy of at least 12 significant digits over the whole range of
parameters specified above.

First, for $x=0$ the integrals $J_l(x)$ take a particularly simple analytic form
\begin{align}
 J_l(0) = 2 - 2\ln2 - H_{l+1/2},
\end{align}
where $H_n$ are the harmonic numbers. This expression can be rewritten as a convenient recursion  $J_{l+1}(0) = J_l(0) -
\frac{2}{2l+3}$, starting with $J_0(0)=0$. The values of $J_l(0)$ constitute an important special case corresponding to
the atomic integrals, but they appear in large numbers also in molecular calculations.

For any value of $x$ the integrals $J_l(x)$ obey the following recursion relation
\begin{align}
 \label{poss}
 J_{l+1}(x) = J_l(x) - 2 F_{l+1}(x),
\end{align}
where 
\begin{align}
 \label{ml}
 \begin{split}
 F_n(T) &= \int_0^1 \text{d} t \, t^{2n}\,e^{-T t^2}. 
 \end{split}
\end{align}
The latter quantity is nothing but the famous Boys function \cite{boys50} considered countless number of times in the
quantum
chemistry literature (see, for example, Refs. \cite{gill91,ishida96,kara10,weiss15} and references therein). Accurate
and efficient methods for calculation of $F_n(T)$ are available and there is no reason for us to elaborate on this
issue.

Returning to the recursion relation (\ref{poss}), its direct use is hampered by a peculiar behaviour of the integrals
$J_l(x)$. Let us temporarily consider $l$ to be a continuous variable. Then, for any fixed $x>0$ the integrals $J_l(x)$
have a root as a function of $l$. For brevity, let us call the exact position of the root (as a function of $x$)
\emph{the critical line}, $l_0(x)$. The exact location of the root cannot be obtained with elementary methods, but we
found that a simple linear function
\begin{align}
\label{crit}
 l_0(x) = 0.44 + 1.17 x
\end{align}
provides a reasonably faithful picture. If the recursion (\ref{poss}) is carried out and the critical
line is crossed, one can expect an unacceptable loss of significant digits due to the cancellations. Therefore, this
simple
approach is inherently numerically unstable, independently of whether the recursion is carried out upward or downward.

One of the possible solutions to this problem is to assert that the critical line is never crossed during the
recursive process. This can be achieved as follows. For an interval of $x$ of approximately unit length we find the
smallest value of $l$ \emph{above} the critical line ($l_a$) and the largest value of $l$ \emph{below} the critical line
($l_b$). Starting from the value at $l_a$ the upward recursion is initiated and
carried out up to the maximal desired value of $l$. Similarly, the downward recursion is initiated at $l_b$ and stopped
at $l=0$. This guarantees that the integrals $J_l(x)$ do not change sign in both sub-recursions and the whole process is
completely numerically stable since the integrals $M_l(x)$ are always positive.

The remaining problem is to evaluate the integrals $J_l(x)$ at $l_a$ and $l_b$ for a given $x$. This is achieved by 
fitting $J_{l_a}(x)$ and $J_{l_b}(x)$ for each interval of $x$. Since the length of each interval is only about unity
the ordinary exponential-polynomial \cite{foot1} fitting is sufficient, i.e.
\begin{align}
 e^{-x}\sum_{k=0}^{N_{fit}} c_k^{(1)} x^k.
\end{align}
The length of the expansion was chosen to be $N_{fit}=11$ in each interval both for $l_a$ and $l_b$.

For $x>x_0\approx 36$ the method described above needs to be slightly modified. This is the point where the critical
line crosses $l=32$. Therefore, for $x>x_0$ all $J_l(x)$ with $l\leq 32$ are positive, and it is sufficient to evaluate
$J_{32}(x)$ by fitting and carry out the recursion (\ref{poss}) downward. We used the following fitting function
\begin{align}
\sum_{k=0}^{N_{fit}'} c_k^{(2)} x^k + \frac{1}{x^{81/2}}\sum_{k=0}^{N_{fit}'} c_k^{(3)} x^k,
\end{align}
with $N_{fit}'=9$. The prefactor in the second term of this expression comes from the asymptotic expansion of $J_l$
which will be introduced in the next paragraph. 

Finally, for $x>x_{\mbox{\scriptsize asym}}$ we use large-$x$ asymptotic expansion of the $J_l(x)$ functions. For $l=0$
the necessary expression was given in Ref. \cite{kpisz09}
\begin{align} 
\label{j0asym}
 J_0(x) = \frac{\sqrt{\pi}}{2x^{3/2}} \sum_{k=0} \frac{(2k+1)!!}{2^k} x^{-k},
\end{align}
and for larger $l$ the corresponding formulae can be obtained by noting that $J_l(x)=-\partial_x J_{l-1}(x)$. The value
of $x_{\mbox{\scriptsize asym}}$ was set to $125$ after some numerical experimentation. Under these conditions the
summation converges to the machine precision after at most $30$ terms. In general, the rate of convergence improves with
increasing $x$ and thus the expansion (\ref{j0asym}) is able to handle arbitrarily large values of
$x>x_{\mbox{\scriptsize asym}}$. Moreover, all terms in Eq. (\ref{j0asym}) are positive and thus no loss 
of digits in the summation is possible. This observation remains valid for $l>0$.

To sum up, the integrals $J_l(x)$ are calculated with a union of three algorithms, involving polynomial fitting,
recursion relations and asymptotic expansion. We note that the efficiency of the resulting code is only somewhat worse
than for the aforementioned Boys function. A \texttt{C++} implementation of the methods described in this section can be
obtained upon request.

\section{Basis set convergence issue}
\label{sec:basis}

Most of the \emph{ab initio} methods used nowadays in the electronic structure theory rely on a basis set for expansion 
of the
exact wavefunction. Consequently, observables obtained with a (necessarily finite) basis set suffer from the basis set
incompleteness error. To allow for a meaningful comparison with the experimental data this error should be estimated and
minimised, if possible. 

One of the prominent techniques applied to remove a bulk fraction of the basis set incompleteness error is the
extrapolation towards the exact theoretical value. However, to ensure that such a procedure is reliable one typically
requires some information on how the calculated values converge towards the exact result as a function of the basis set
size. For example, it was shown by Hill \cite{hill85} that the nonrelativistic energy converges as $L^{-3}$, where $L$
is the largest angular momentum present in the basis set. It can be shown that some relativistic corrections converge
even slower, as $L^{-1}$. This was numerically observed in Refs. \cite{ottschofski97,halkier00} and later proved by
Kutzelnigg \cite{kutz08}. In this case the values calculated with a finite basis set can be in error of tens of percents
and extrapolation is necessary to arrive at a reliable result.

Concerning the Araki-Sucher correction, it has never been assessed thus far how the results obtained with finite basis
sets converge as a function of the largest angular momentum included. To answer this question we consider the ground
state of the helium atom as a model system where a strict asymptotic result can be obtained. Further in the paper we
show numerically that the main conclusions are valid also for many-electron many-centre systems. This allows for a
reliable extrapolation towards the complete basis set limit, dramatically improving the final results.

\subsection{Definitions and notation}

We consider the ground 1$^1$S state of the helium atom with the exact wavefunction given by $\Psi(\vecr_1, \vecr_2)$,
where $\vecr_i$ are the positions of the electrons and $\vecr_{12} = \vecr_1 - \vecr_2$. The corresponding wavefunction
can be represented as
\begin{align}
\label{pwe1}
\Psi(\vecr_1, \vecr_2) = \sum_{l=0}^\infty \Psi_l(r_1,r_2) P_l(\cos \theta_{12}),
\end{align}
where $r_i = |\vecr_i|$, $P_l$ are the Legendre polynomials, and $\theta_{12}$ is the angle between vectors $\vecr_1$,
$\vecr_2$. The above expression is dubbed \emph{the partial wave expansion} (PWE) by many authors and we shall follow
this nomenclature for the wavefunctions and operators.

It is natural to define a family of approximants to the exact wavefunction by truncating Eq. (\ref{pwe1}) at a given
$L$, i.e.
\begin{align}
\label{pwe2}
\Psi_L(\vecr_1, \vecr_2) = \sum_{l=0}^L \Psi_l(r_1,r_2) P_l(\cos \theta_{12}).
\end{align}
The Araki-Sucher correction can be than approximated as the $a\rightarrow0$ limit of the following expectation values
\begin{align}
\label{exp1}
\langle\hat{P}_a (r_{ij}^{-3})\rangle_L = \langle \Psi_L|\hat{P}_a (r_{ij}^{-3})| \Psi_L\rangle / \langle \Psi_L| \Psi_L\rangle.
\end{align}
Obviously, in the infinite $L$ limit this series converges to the exact value, $\langle\hat{P}_a (r_{ij}^{-3})\rangle$.
Therefore, we may consider the error
\begin{align}
\label{err}
\epsilon_L(a) = \langle\hat{P}_a (r_{ij}^{-3})\rangle - \langle\hat{P}_a (r_{ij}^{-3})\rangle_L
\end{align}
as a function of $L$ and ask what is the asymptotic form of $\epsilon_L(a)$ at large $L$. After taking the
$a\rightarrow0$ limit one recovers the actual result for the Araki-Sucher correction. This is only a precise
mathematical restatement of the intuitive picture presented at the beginning of this section.

All derivations presented further rely on the seminal work of Hill and the methods introduced therein \cite{hill85}. The
original presentation of Hill relies on a chain of postulates which are extremely difficult to prove strictly but are
nonetheless very physically sound and hard to deny (especially in the face of ample numerical evidence). First, the
denominator in Eq. (\ref{exp1}) can be replaced by unity as it converges much faster than the numerator and does not
contribute in the leading order. Second, for large $L$ the dominant contribution to the integral in Eq. (\ref{exp1})
comes from the region around the electrons coalescence points. The famous Kato cusp condition \cite{kato57} teaches us
that in this regime the exact wavefunction behaves as
\begin{align}
\label{kato}
 \Psi(\vecr_1, \vecr_2) = \Psi(r,r,0) \left( 1 + \half r_{12} \right) + \mathcal{O}\big(r_{12}^2\big),
\end{align}
where $\Psi(r,r,0)$ is the value of the exact wavefunction at $r_{12}=0$. By using these assumptions modulus square of
the wavefunction present is rewritten as
\begin{align}
\label{mod2}
|\Psi(\vecr_1, \vecr_2)|^2 = |\Psi(r,r,0)|^2 \left( 1 + r_{12} \right) + \mathcal{O}\big(r_{12}^2\big).
\end{align}
Finally, let us recall PWE for $r_{12}$
\begin{align}
\label{pwer12}
 r_{12} &= \sum_{l=0}^\infty \{ r_{12} \}_l \, P_l(\cos \theta_{12}),\\
 \{ r_{12} \}_l &= \frac{1}{2l+3} \frac{\rs^{l+2}}{\rb^{l+1}} - \frac{1}{2l-1} \frac{\rs^l}{\rb^{l-1}},
\end{align}
where $\rs=\min(r_1,r_2)$, $\rb=\max(r_1,r_2)$. This expression is closely related to the well-known Laplace expansion
of the potential.

\subsection{PWE for the Araki-Sucher distribution}

Throughout the presentation we shall need PWE for the distribution of Eq. (\ref{AS2}),
\begin{align}
\label{pweas1}
 \widehat{P}_a\left(r_{12}^{-3}\right) = \sum_{l=0}^\infty \mathcal{A}_l(r_1,r_2;a)\,P_l(\cos \theta_{12}),
\end{align}
where the radial coefficients are defined formally through the expression
\begin{align}
\label{pweas2}
\begin{split}
&\mathcal{A}_n(r_1,r_2;a) = \frac{2n+1}{2} \\
 &\times\int_0^\pi d\theta_{12}\,\sin \theta_{12}\,\widehat{P}_a\left(r_{12}^{-3}\right)P_n(\cos \theta_{12}). 
\end{split}
\end{align}
Derivation of the explicit expression for $\mathcal{A}_l(r_1,r_2;a)$ is fairly straightforward and relies solely on Eq.
(\ref{pweas2}). However, it requires some tedious technical algebra. In order to shorten the main article, we decided to
move the entire derivation to Appendix \ref{app:appa}. Herein, we present only the final result
\begin{align}
\label{ap}
\begin{split}
 \mathcal{A}_n(r_1,r_2;a) &= \mathcal{A}_n'(r_1,r_2;a) + \mathcal{A}_n''(r_1,r_2;a) \\
 &+ \mathcal{A}_n'''(r_1,r_2;a) + \mathcal{O}\left(a\right),
\end{split}
\end{align}
where
\begin{align}
 \label{ap1}
 \mathcal{A}_n'(r_1,r_2;a) = \theta\left(|r_1-r_2|-a\right)\frac{\left(2n+1\right)\rs^n\rb^{-n-1}}{\rb^2-\rs^2},
\end{align}
\begin{align}
 \label{ap2}
 \begin{split}
 \mathcal{A}_n''(r_1,r_2;a) &= \theta\left(a-|r_1-r_2|\right)\theta\left(r_1+r_2-a\right) \\ 
 &\times \frac{2n+1}{2a} \frac{1}{r_1r_2}P_n\left(\frac{r_1^2+r_2^2}{2r_1r_2}\right),
 \end{split}
\end{align}
and
\begin{align}
 \label{ap3}
 \mathcal{A}_n'''(r_1,r_2;a) = \big(2n+1\big)\left(\gamma +\ln a\right) r_1^{-2}\, \delta(r_1-r_2). 
\end{align}
Note that only the leading-order terms in $a$ have been retained in the above formulae. This is justified because the
actual Araki-Sucher correction of Eq. (\ref{AS1}) involves the $a\rightarrow0$ limit and all higher-order contributions
in $a$ vanish.

\subsection{Large-$L$ asymptotic formula for $\epsilon_L(a)$}

Let us insert Eqs. (\ref{mod2}), (\ref{pwer12}) and (\ref{pweas1}) into Eq. (\ref{exp1}) and change the variables to
$r_1$, $r_2$, $\theta_{12}$. Integration over the remaining 3 variables gives $8\pi^2$ and integration over
$\theta_{12}$ is trivial due to the orthogonality of the Legendre polynomials. The error is given by
\begin{align}
\label{el}
\begin{split}
 \epsilon_L(a) = 16\pi^2 \sum_{l=L+1}^\infty \frac{1}{2l+1}\int_0^\infty dr_1\int_0^\infty dr_2  \\
 \times r_1^2\,r_2^2\, |\Psi(r,r,0)|^2\, \{ r_{12} \}_l\, \mathcal{A}_l(r_1,r_2;a),
\end{split}
\end{align}
where the factor $r_1^2\,r_2^2$ comes from the volume element. Let us define the following quantities
\begin{align}
\label{i1}
\begin{split}
 I_n'(a) &= \frac{1}{2n+1}\int_0^\infty dr_1\int_0^\infty dr_2\, r_1^2\,r_2^2  \\
 &\times |\Psi(r,r,0)|^2\, \{ r_{12} \}_n\, \mathcal{A}_n'(r_1,r_2;a),
\end{split}
\end{align}
which are natural constituents of Eq. (\ref{el}). Analogous definitions hold for the doubly primed and triply primed
quantities in accordance with Eqs. (\ref{ap})-(\ref{ap3}) and for the sum of the three (without the prime).

Starting with Eq. (\ref{i1}), we change the variables to $\rb$ and $\rs$, insert the explicit form of Eqs.
(\ref{pwer12}) and (\ref{ap1}), and execute the Heaviside theta to arrive at the result
\begin{align}
 \begin{split}
  I_n'(a) &= 2\int_0^\infty d\rb |\Psi(\rb,\rb,0)|^2 \int_0^{\rb-a} d\rs \\ 
  &\times \frac{1}{\rb^2-\rs^2}\left[ \frac{\rs^{2n+4}\,\rb^{-2n}}{2n+3}-\frac{\rs^{2n+2}\,\rb^{2n-2}}{2n-1} \right].
 \end{split}
\end{align}
The inner integral can be brought into a closed form, but it is much simpler to obtain the leading-order expression in
$a$ from the integration by parts. This leads to
\begin{align}
\label{ipp1}
\begin{split}
 &I_n'(a) = \frac{4}{(2n-1)(2n+3)} \int_0^\infty d\rb\,\rb^3 \\
 &\times |\Psi(\rb,\rb,0)|^2 \left[ \gamma - \half + \ln \frac{(2n+3)a}{\rb} \right] + \mathcal{O}\left(a\right),
\end{split}
\end{align}
where additionally some higher-order terms in $1/n$ have been neglected.

Passing to the doubly primed quantities, we insert Eq. (\ref{ap2}) into Eq. (\ref{i1}) and after elementary
rearrangements and a change of variable, we obtain
\begin{align}
\label{ip2}
 \begin{split}
 &I_n''(a) = \frac{1}{a} \int_0^\infty d\rb\int_0^{\rb} d\rs\,|\Psi(\rb,\rb,0)|^2 \\
 &\times\, \theta(a-\rb+\rs)\,\theta(\rb+\rs-a) \\
 &\times P_n\left(\frac{\rb^2+\rs^2}{2\rb\rs}\right)\left[ \frac{\rs^{n+3}\,\rb^{-n}}{2n+3}-\frac{\rs^{n+1}\,\rb^{n-2}}{2n-1} \right].
 \end{split}
\end{align}
By the virtues of the $\theta$ function the integral can be rewritten to the form
\begin{align}
\begin{split}
 &\frac{1}{a} \int_0^\infty d\rb\int_0^{\rb} d\rs\, \theta(a-\rb+\rs)\\
 &\times \theta(\rb+\rs-a)\,\ldots = \frac{1}{a}\bigg[ \int_{a/2}^a d\rb \int_{a-\rb}^{\rb} d\rs \\
 &+ \int_a^\infty d\rb \int_{\rb-a}^{\rb}d\rs\,\ldots\bigg]
\end{split}
\end{align}
It turns out that in our case the first integral gives zero contribution (in the small $a$ limit). The inner integral of
the second component can be expanded as a powers series in $a$. The first term vanishes and only the second (i.e.,
proportional to $a$) has to be retained giving
\begin{align}
\begin{split}
&\int_{\rb-a}^{\rb}d\rs\, P_n\left(\frac{\rb^2+\rs^2}{2\rb\rs}\right)\left[ \frac{\rs^{n+3}\,\rb^{-n}}{2n+3}-\frac{\rs^{n+1}\,\rb^{n-2}}{2n-1} \right] = \\
&a\cdot P_n\left(\frac{\rb^2+\rs^2}{2\rb\rs}\right)\left[ \frac{\rs^{n+3}\,\rb^{-n}}{2n+3}-\frac{\rs^{n+1}\,\rb^{n-2}}{2n-1}\right]\bigg|_{\rs=\,\rb} = \\
&a\cdot \frac{-4\,\rb^3}{(2n-1)(2n+3)} + \mathcal{O}\left(a^2\right).
\end{split}
\end{align}
Upon reinserting into Eq. (\ref{ip2}) and rearranging one obtains
\begin{align}
\begin{split}
 I_n''(a) &= \frac{-4}{(2n-1)(2n+3)} \int_0^\infty d\rb\,\rb^3 \\
 &\times |\Psi(\rb,\rb,0)|^2 + \mathcal{O}\left(a\right).
\end{split}
\end{align}
The last integral $I_n'''(a)$ is the simplest to evaluate. One inserts Eq. (\ref{ap3}) into Eq. (\ref{i1}) and executes the Dirac delta to arrive at
\begin{align}
 \begin{split}
 I_n'''(a) &= -\frac{4(\gamma+\ln a)}{(2n-1)(2n+3)} \\
 &\times\int_0^\infty d\rb\,\rb^3\,|\Psi(\rb,\rb,0)|^2, 
 \end{split}
\end{align}
without invoking any approximations. Finally, we add up the three integrals evaluated above
\begin{align}
\begin{split}
\label{finali}
 &I_n(a) = \frac{1}{4\pi^2} \frac{1}{(2n-1)(2n+3)} \times\\
 & \bigg[ \left( \ln \left(2n+3\right) - \frac{3}{2}\right) \mathcal{J}_3 - \mathcal{J}_{\ln} \bigg] + \mathcal{O}\left(a\right),
\end{split}
\end{align}
where
\begin{align}
\label{j3}
 &\mathcal{J}_3 = 16\pi^2 \int_0^\infty dr\,r^3\,|\Psi(r,r,0)|^2,\\
 \label{jln}
 & \mathcal{J}_{\ln} = 16\pi^2 \int_0^\infty dr\,r^3\,\ln r\,|\Psi(r,r,0)|^2.
\end{align}
One can see that in the final expression all logarithmic singularities cancel out. Therefore, we can now take the limit $a\rightarrow0$ removing all
higher-order terms in $a$.

Let us now return to the formula for the error, Eq. (\ref{el}) at $a=0$. Making use of Eq. (\ref{finali}) and after some algebra the result
can be written as
\begin{align}
 \begin{split}
 &\epsilon_L(0) = 4 \mathcal{J}_3 \sum_{n=L+1}^\infty \frac{\ln\left(2n+3\right)}{(2n-1)(2n+3)} \\ 
 &- 4\left(\frac{3}{2}\mathcal{J}_3 + \mathcal{J}_{\ln}\right) \sum_{n=L+1}^\infty \frac{1}{(2n-1)(2n+3)}.
 \end{split}
\end{align}
The first infinite sum is nontrivial to evaluate, but we can utilise the Euler-Maclaurin resummation formula to get the
large-$L$ asymptotics. 
This gives the leading-order expressions and their error estimates
\begin{align}
 &\sum_{n=L+1}^\infty \frac{\ln\left(2n+3\right)}{(2n-1)(2n+3)} = \frac{1+\ln 2L}{4L} + \mathcal{O}\left(\frac{\ln L}{L^2}\right),\\
 &\sum_{n=L+1}^\infty \frac{1}{(2n-1)(2n+3)} = \frac{1}{4L} + \mathcal{O}\left(L^{-2}\right).
\end{align}
Finally, we rewrite the error formula as
\begin{align}
\label{final}
 \epsilon_L(0) = \mathcal{J}_3 \frac{\ln 2L}{L} - \frac{1}{L}\Big( \mathcal{J}_{\ln} + \half \mathcal{J}_3\Big) + \mathcal{O}\left(\frac{\ln L}{L^2}\right),
\end{align}
which indicates a very slow, i.e., logarithmic convergence of the Araki-Sucher correction towards the exact
value. In fact, the convergence rate is even slower than for the aforementioned relativistic corrections \cite{kutz08}.
Nevertheless, the above formula gives precise information on how the values from the finite basis sets should be
extrapolated.

\section{Numerical results}
\label{sec:numer}

\begin{table}[t]
\caption{Total electronic energies and the expectation values of the Araki-Sucher distribution for the helium atom.
Extrapolations were performed with help of Eq. (\ref{final}) in case of the Araki-Sucher correction and with the
standard $X^{-3}$ formula in case of the energy \cite{hill85}. All values are given in the atomic units.}
\label{tabhelium}
\begin{ruledtabular}
\begin{tabular}{lcc}
 basis & $-E$ & $\langle\widehat{P}\left(r_{12}^{-3}\right)\rangle$ \\[0.6ex]
\hline \\[-2.2ex]
 d3Z & 2.90 170 & 0.470 \\
 d4Z & 2.90 285 & 0.541 \\
 d5Z & 2.90 328 & 0.595 \\
 d6Z & 2.90 347 & 0.637 \\
 d7Z & 2.90 356 & 0.670 \\
\hline \\[-2.2ex]
extrapolation & 2.90 372 & 1.003 \\
\hline \\[-2.2ex]
reference$^{\mbox{\scriptsize a}}$ & 2.90 372 438 & 0.989 274 \\
\end{tabular}
\begin{flushleft}\vspace{-0.2cm}
$^{\mbox{\scriptsize a}}${\small from the paper of Frolov \cite{frolov15}}; all digits shown are correct\\
\end{flushleft}
\end{ruledtabular}
\end{table}

\begin{table}[t]
\caption{Expectation values of the Araki-Sucher distribution for the hydrogen molecule ($R$ denotes the internuclear
distance). Reference values are given in the third and the fourth column for comparison purposes. All values are given
in the atomic units.}
\label{tabh2}
\begin{ruledtabular}
\begin{tabular}{cccc}
 $R$ & \multicolumn{3}{c}{$\langle\widehat{P}\left(r_{12}^{-3}\right)\rangle$} \\[0.6ex]
  & this work & Ref. \cite{stanke17} & Ref. \cite{kpisz09} \\[0.6ex]
\hline \\[-2.2ex]
0.1 & 0.8742 & 0.8707 & 0.8847 \\ 
0.6 & 0.8042 & 0.7782 & 0.7775 \\ 
0.8 & 0.6857 & 0.6698 & 0.6696 \\ 
1.0 & 0.6014 & 0.5714 & 0.5712 \\ 
1.4 & 0.4305 & 0.4135 & 0.4143 \\ 
1.7 & 0.3356 & 0.3248 & 0.3250 \\ 
2.0 & 0.2542 & 0.2550 & 0.2554 \\ 
2.6 & 0.1531 & 0.1535 & 0.1555 \\ 
6.0 & 0.0060 & 0.0025 & 0.0063 \\
\end{tabular}
\end{ruledtabular}
\end{table}

\subsection{Benchmark calculations}

To verify that the method of calculation of the matrix elements of the Araki-Sucher distribution and the extrapolation
scheme (\ref{final}) are both valid we performed calculations for several systems where reference values of this
quantity are known to a sufficient accuracy. The includes the helium atom (He), lithium atom (Li) and its cation
(Li$^+$), beryllium atom (Be) and its cation (Be$^+$), and the hydrogen molecule (H$_2$). Expectation values of the
Araki-Sucher distribution were computed by using the finite-field approach. Suitable values of the displacement
parameter were found individually for each system by trial-and-error. Typically, a value of about $10^{-5}$ was
optimal. For the two- and three-electron systems (He, Li$^+$, H$_2$, Li, Be$^+$) we used the full CI method to solve the
electronic Schr\"{o}dinger equation (this method is exact in the complete basis set limit). For larger systems we
employed the CCSD(T) method \cite{ragha89,piecuch02}. All electronic structure calculations reported in this work were
performed with help of a locally modified version of the \textsc{Gamess} program package \cite{gamess1,gamess2}. For the
helium atom we used the customised basis sets developed by Cencek et al. \cite{cencek12}. For the hydrogen molecule,
lithium and beryllium (both neutral atoms and cations) the standard basis sets developed in Refs.
\cite{dunning89,prascher11} were employed.

In Table \ref{tabhelium} we show results for the calculations for the helium atom. One can see a very slow convergence
of the results with the size of the basis set. To overcome this difficulty we applied a two-point extrapolation
formula, Eq. (\ref{final}). Note that in the present case we do not extrapolate with respect to the
maximal angular momentum present in the basis ($L$) but rather with respect to the so-called cardinal number ($X$)
\cite{dunning89}. This does not change the asymptotic formula (\ref{final}) but changes values of the numerical
coefficients in the expansion. Therefore, we do not attempt to compare the values obtained by fitting with the analytic
results given by Eqs. (\ref{j3}) and (\ref{jln}). Nonetheless, the quality of the extrapolation is very good.
Extrapolation from the basis sets $X=3,\ldots,6$ reduces the error from about 30\% to less than 1.5\% (cf. Table
\ref{tabhelium}). One can safely say that the extrapolation is mandatory to obtain results of any reasonable quality.

Let us now pass to the calculations for the hydrogen molecule, H$_2$. We calculated the
Araki-Sucher correction for several internuclear distances and compared them with more accurate values given by
Piszczatowski et al. \cite{kpisz09} and by Stanke et al. \cite{stanke17} obtained with the explicitly correlated
Gaussian wavefunctions. Our extrapolated results are given in Table \ref{tabh2} and compared with the two sets of
reference values. One can see a reasonable agreement between the present results and Refs. \cite{stanke17,kpisz09}. The
biggest absolute deviation from the values of Piszczatowski et al. \cite{kpisz09} is about 4\%. This is only slightly
larger than for the helium atom. This error increase can be (at least partially) attributed to the fact that a larger
d7Z basis set was used for helium atom while calculations for the hydrogen molecule were restricted to the d6Z basis.

\begin{table}[t]
\caption{Expectation values of the Araki-Sucher distribution for the lithium and beryllium atoms (Li, Be),
and their cations (Li$^+$, Be$^+$). All values are given in the atomic units.}
\label{tablibe}
\begin{ruledtabular}
\begin{tabular}{lllll}
 basis & \multicolumn{4}{c}{$\langle\widehat{P}\left(r_{12}^{-3}\right)\rangle$} \\[0.6ex]
       & \multicolumn{1}{c}{Li} & \multicolumn{1}{c}{Li$^+$} & \multicolumn{1}{c}{Be} & \multicolumn{1}{c}{Be$^+$}
\\[0.6ex]
\hline \\[-2.2ex]
 d3Z & $-$2.891 & $-$2.924 & $-$15.32 & $-$15.38 \\
 d4Z & $-$2.194 & $-$2.241 & $-$13.52 & $-$13.61 \\
 d5Z & $-$1.718 & $-$1.774 & $-$12.06 & $-$12.18 \\
 d6Z & $-$1.397 & $-$1.460 & $-$11.25 & $-$11.38 \\
\hline \\[-2.2ex]
extrapolation & $+$0.267 & $+$0.173 & $-$7.320 & $-$7.505 \\
\hline \\[-2.2ex]
reference & $+$0.2734$^{\mbox{\scriptsize a}}$ & $+$0.1789$^{\mbox{\scriptsize b}}$ & $-$7.3267$^{\mbox{\scriptsize c}}$
& $-$7.5146$^{\mbox{\scriptsize a}}$ \\
\end{tabular}
\begin{flushleft}\vspace{-0.2cm}
$^{\mbox{\scriptsize a}}${\small Ref. \cite{puchalski08}}\;\;
$^{\mbox{\scriptsize b}}${\small Ref. \cite{frolov14}}\;\;
$^{\mbox{\scriptsize c}}${\small Ref. \cite{pachucki04b}}
\end{flushleft}
\end{ruledtabular}
\end{table}

Finally, in Table \ref{tablibe} we show results of the calculations for the lithium and beryllium atoms as well as the
corresponding cations. Our extrapolated values are compared with the reference data taken from the papers of Frolov et
al. \cite{frolov15,frolov14} and Pachucki et al. \cite{pachucki04b,puchalski08}. The errors are consistently within the
range of 1$-$2\%. Only for the lithium cation the accuracy is slightly worse ($\approx\,$3\%) but this is probably
accidental. We can also check how well relative differences are reproduced in our method. To this end, we calculate 
contributions of the Araki-Sucher term to the ionisation energies of the lithium and beryllium atoms and compare the
results with Refs. \cite{pachucki04b,puchalski08}. In both cases we find a remarkable agreement within approx. 1\% of
the total value.

To sum up, the method of calculating the Araki-Sucher correction proposed here is fundamentally valid
and useful in practice. By comparing our results with the reference data available in the literature for several
few-body systems we conclude that it is capable of reaching an accuracy of a few percents or better. This is
true provided that sufficiently large basis sets and accurate electronic structure methods are employed. Moreover,
extrapolation to the complete basis set limit must be performed in every case. The theoretically derived leading-order
formula (\ref{final}) is very efficient in this respect.

\subsection{Results for many-electron systems}

\begin{table}[t]
\caption{Expectation values of the Araki-Sucher distribution for the magnesium atom (Mg) and its cation (Mg$^+$), and
the argon atom (Ar) and dimer (Ar$_2$). All values are given in the atomic units. The uncertainties in the final values
are estimated to be about 5\%.}
\label{tablmany}
\begin{ruledtabular}
\begin{tabular}{lllll}
 basis & \multicolumn{4}{c}{$\langle\widehat{P}\left(r_{12}^{-3}\right)\rangle$} \\[0.6ex]
       & \multicolumn{1}{c}{Mg} & \multicolumn{1}{c}{Mg$^+$} & \multicolumn{1}{c}{Ar} & \multicolumn{1}{c}{Ar$_2$}
\\[0.6ex]
\hline \\[-2.2ex]
 d2Z & $-$1371.3 & $-$1370.5 & $-$6090.3 & $-$12179.9 \\
 d3Z & $-$1331.1 & $-$1330.2 & $-$6022.7 & $-$12044.8 \\
 d4Z & $-$1315.9 & $-$1315.0 & $-$5963.4 & $-$11926.1 \\
 d5Z & $-$1295.0 & $-$1294.2 & $-$5917.1 & $-$11833.7 \\
\hline \\[-2.2ex]
extrapolation & $-$1220.7 & $-$1221.4 & $-$5440.8 & $-$10881.0 \\
\end{tabular}
\end{ruledtabular}
\end{table}

The biggest advantage of the method proposed here is that it can be applied to systems much larger than studied
previously. This includes not only many-electron atoms, but also diatomic and even polyatomic molecules. To
illustrate this we performed calculations for several many-body systems - the magnesium atom (Mg) and its ion (Mg$^+$)
and the argon atom (Ar) and its dimer (Ar$_2$). In the case of Mg and Mg$^+$ we employed the IP-EOM-CCSD-3A method
\cite{gour05,gour06} and the basis sets ``aug-cc-pwCVX`` reported in Ref. \cite{prascher11}. For the Ar and Ar$_2$
systems we used the CCSD(T) method and the basis sets ``disp-XZ+2/AE'' developed by Patkowski and Szalewicz
\cite{patkowski10} specifically for accurate description of the argon dimer. The results are shown in Table
\ref{tablmany}. Overall, the rate of convergence of the values obtained in finite basis sets is similar as for the
helium atom which validates the extrapolation formula (\ref{final}) for many-electron systems. We can estimate that the
accuracy of the results shown in Table \ref{tablmany} is not worse than 5\%.

With the help of the results from Table \ref{tablmany} one can also calculate the contribution of the Araki-Sucher
correction to the ionisation energy of the magnesium atom and interaction energy the argon dimer. The former quantity
is approximately equal to $-$0.02 cm$^{-1}$ (the negative sign indicates that this correction decreases the ionisation
energy). While this value seems to be very small we note that it is of the same order of magnitude as the present-day
experimental uncertainty in the measurement of the ionisation energy of the magnesium atom, 0.03 cm$^{-1}$
\cite{martin80,chang87,kaufman91}. For the argon dimer we calculate that the contribution to the interaction energy of
the Araki-Sucher term is equal to 0.02 cm$^{-1}$. Again, this value has to be put into context. The total interaction
energy of the argon dimer is approximately 99 cm$^{-1}$. Therefore, while the Araki-Sucher contribution is small on the
absolute scale, it becomes non-negligible in relation to other subtle effects. Moreover, the theoretical accuracy
attainable for the argon dimer at present \cite{patkowski10,slavicek03,patkowski05} is already quite close to the level
where the QED effects come into play.

\section{Conclusions}
\label{sec:concl}

In the present work we have put forward a general scheme to calculate the Araki-Sucher correction for many-electron
systems. Several obstacles had to be removed to accomplish this goal. First, the complicated two-electron integrals
involving the Araki-Sucher distribution have been solved with help of the McMurchie-Davidson technique (within the
Gaussian-type orbitals basis set). It has been shown that they can be expressed through a family of one-dimensional
integrals. Recursive and numerically stable computation of the latter integrals has been discussed in details.

Second, the issue of convergence of the results with respect to the size of the basis set has been considered.
We have demonstrated a slow convergence pattern ($\ln 2L/L$ in the leading order) towards the complete basis set limit.
This result has been verified by comparing with reference data for the helium atom. With the analytic information about
the convergence at hand, extrapolations have been used to improve the accuracy of the results. The accuracy of about 1\%
has been achieved in this case.

To confirm the validity of the proposed approach we have performed calculations for several few- and many-electron
systems. First, we have concentrated on small systems (e.g., few-electron atoms, hydrogen molecule) for which accurate
reference values are available in the literature. A consistent accuracy of few percents have been obtained and the
molecular results are only slightly less accurate than the atoms. Next, we have moved on to many-electron systems. We
have estimated the contribution of the Araki-Sucher correction to the ionisation energy of the magnesium atom and
interaction energy of the argon dimer. The final values of the Araki-Sucher correction are comparable to the
present day experimental uncertainties of the measurements.

\begin{acknowledgments}
This work was supported by the Polish National Science Centre through the project 2016/21/B/ST4/03877. The authors would
like to thank Bogumi\l~Jeziorski for reading and commenting on the manuscript.
\end{acknowledgments}

\appendix
\section{Partial wave expansion of the Araki-Sucher distribution}
\label{app:appa}

In this section we present details of the derivation of Eqs. (\ref{pweas1})-(\ref{ap3}). Let us start
with the definition (\ref{pweas1}) and split it into two parts, $\mathcal{A}_n(r_1,r_2;a) =
\mathcal{A}^{(1)}_n(r_1,r_2;a) + \mathcal{A}^{(2)}_n(r_1,r_2;a)$, with
\begin{align}
\label{a1}
\begin{split}
 \mathcal{A}^{(1)}_n(r_1,r_2;a) &=
 \frac{2n+1}{2} \int_0^\pi d\theta_{12}\,\sin \theta_{12}\, \\
 &\times \frac{\theta \left(r_{12}-a\right)}{r_{12}^3} P_n(\cos
\theta_{12}),
\end{split}
\end{align}
and
\begin{align}
\label{a2}
\begin{split}
 &\mathcal{A}^{(2)}_n(r_1,r_2;a) = 2\pi\big(2n+1\big)\left(\gamma +\ln a\right) \\
 &\times\int_0^\pi d\theta_{12}\,\sin \theta_{12}\, \delta (\vecr_{12}) P_n(\cos \theta_{12}).
\end{split}
\end{align}
The second of these integrals is straightforward to evaluate because in the present context
\begin{align}
\delta (\vecr_{12}) = \frac{\delta(r_1-r_2)}{2\pi r_1^2} \frac{\delta(\theta_{12})}{\sin \theta_{12}}.
\end{align}
Upon inserting back into Eq. (\ref{a2}) the integration over the angle becomes straightforward and with help of the
expression $P_n(1)=1$ one arrives at
\begin{align}
\label{a2f}
 \mathcal{A}^{(2)}_n(r_1,r_2;a) = \big(2n+1\big)\left(\gamma +\ln a\right) r_1^{-2}\, \delta(r_1-r_2).
\end{align}
Evaluation of the first term $\mathcal{A}^{(1)}_n(r_1,r_2;a)$ is much more complicated. Changing the integration
variable in Eq. (\ref{a1}) to $r_{12} = \big( r_1^2 + r_2^2 - 2 r_1 r_2 \cos \theta_{12}\big)^{1/2}$ gives
\begin{align}
 \begin{split}
 \mathcal{A}_n(r_1,r_2;a) &= \frac{2n+1}{2r_1r_2} \int_{|r_1-r_2|}^{r_1+r_2} dr_{12}\,
 \theta\left(r_{12}-a\right)\\
 &\times\frac{1}{r_{12}^2} P_n\left( \frac{r_1^2+r_2^2-r_{12}^2}{2r_1r_2} \right).
 \end{split}
\end{align}
To get rid of the theta function under the integral sign we need to distinguish three possible (and disjoint) cases.
First, assuming that $a<|r_1-r_2|$ the integration range remains unchanged because $\theta \left(r_{12}-a\right)$ is
equal to the unity there. The second case is $|r_1-r_2|<a<r_1+r_2$ - the integrand vanishes whenever $r_{12} < a$ so
that the lower integration limit has to be shifted to $a$. The third
case is $a>r_1+r_2$ - the result is zero because the integrand vanishes here. With this reasoning the integral can be
rewritten as
\begin{align}
\label{a21}
\begin{split}
 &\mathcal{A}_n(r_1,r_2;a) = \theta\left(|r_1-r_2|-a\right)\frac{2n+1}{2} \\
 &\times\int_{-1}^{+1}du\, \big(r_1^2+r_2^2-2r_1r_2u\big)^{-3/2}P_n(u) \\
 &+ \theta\left(a-|r_1-r_2|\right)\theta\left(r_1+r_2-a\right)\frac{2n+1}{2} \\
 &\times\int_{-1}^{u(a)}du\, \big(r_1^2+r_2^2-2r_1r_2u\big)^{-3/2}P_n(u)
\end{split}
\end{align}
after the change of variables to $u = (r_1^2+r_2^2-r_{12}^2)/2r_1r_2$ and with the shorthand notation $u(a) =
(r_1^2+r_2^2-a^2)/2r_1r_2$. The first integral is evaluated with elementary methods
\begin{align}
\label{a22}
 \begin{split}
  \int_{-1}^{+1}du\, \big(r_1^2+r_2^2-2r_1r_2u\big)^{-3/2}P_n(u) = \\
  \frac{2\rs^n\rb^{-n-1}}{(\rb-\rs)(\rb+\rs)}.
 \end{split}
\end{align}
The second integral is more complicated because of the function in the upper integration limit. It is probably quite
difficult to derive the explicit expression for this integral but fortunately we require only the leading-order term in
$a$. The higher-order terms vanish in the final result due to the $a\rightarrow0$ limit. To extract the leading-order
contribution we return to the original variable and integrate by parts once
\begin{align}
\label{a23}
\begin{split}
 &\int_{-1}^{u(a)}du\, \big(r_1^2+r_2^2-2r_1r_2u\big)^{-3/2}P_n(u) = \\
 &\frac{1}{r_1r_2} \int_a^{r_1+r_2} \frac{dt}{t^2}\, P_n\left(\frac{r_1^2+r_2^2-t^2}{2r_1r_2}\right) = \\
 &-\frac{(-1)^n}{r_1r_2(r_1+r_2)} + \frac{1}{a}\frac{1}{r_1r_2}P_n\left(\frac{r_1^2+r_2^2-a^2}{2r_1r_2}\right) \\
 &- \frac{1}{r_1^2r_2^2} \int_a^{r_1+r_2} dt\,P_n'\left(\frac{r_1^2+r_2^2-t^2}{2r_1r_2}\right).
\end{split}
\end{align}
The first and the second terms are of the order $a^0$ and $a^{-1}$, respectively. By integrating by parts
again one can show that the last term is also of the order $a^0$. Therefore, we can write
\begin{align}
\label{a24}
 \begin{split}
  &\int_{-1}^{u(a)}du\, \big(r_1^2+r_2^2-2r_1r_2u\big)^{-3/2}P_n(u) = \\
  &\frac{1}{a}\frac{1}{r_1r_2}P_n\left(\frac{r_1^2+r_2^2}{2r_1r_2}\right) + \mathcal{O}\left(a^0\right),
 \end{split}
\end{align}
which is sufficient for the present purposes. Finally, to arrive at Eqs. (\ref{ap})-(\ref{ap3}) from the main text one
has to gather Eqs. (\ref{a2f}), (\ref{a21}), (\ref{a22}), (\ref{a23}), and (\ref{a24}) and rearrange.


\begin{thebibliography}{000}
\bibitem{foot1} We investigated the possibility of using the Tchebyshev polynomials
\end{thebibliography}
\end{document}